\documentclass[11pt,final]{article}
\usepackage[color,final]{showkeys}

\usepackage{version}
\usepackage{amsmath}
\usepackage{graphicx}
\usepackage{amsthm}
\usepackage{amssymb}
\usepackage{alltt}
\usepackage{color}
\usepackage{epic}
\usepackage{eepic}
\usepackage{fancyheadings}
\usepackage{shadow}
\usepackage{array}
\usepackage{caption2}
\usepackage{enumerate}
\usepackage{ulem}
\usepackage{epsfig}
\usepackage{multicol}
\usepackage{ifthen}
\usepackage{bbold}

\newcommand{\brk}[1]{\left(#1\right)}          
\newcommand{\Brk}[1]{\left[#1\right]}          
\newcommand{\BRK}[1]{\left\{#1\right\}}        
\newcommand{\brkco}[1]{\left[#1\right)}       

\newcommand{\goto}{\rightarrow}

 {

\newtheorem*{prop*}{Proposition}
 }

\newcounter{sect}

\newcommand{\oneover}[1]{{\frac{1}{#1}}}
\newcommand{\half}{\frac{1}{2}}

\newcommand{\quarter}{\oneover{4}}

\newcommand{\one}{\mathbb 1}

\newcommand{\LL}{{\cal L}}

\renewcommand{\phi}{\varphi}

\newcommand{\eps}{\varepsilon}

\newcommand{\MATLAB}{M\textsc{atlab}\ }

\title{The Optimal Shape of a Javelin}
\author{Yossi Farjoun\footnote{Department of Mathematics, UC
    Berkeley, Berkeley CA; e-mail: yfarjoun@math.berkeley.edu}\and
  John Neu\footnote{Department of Mathematics, UC Berkeley, Berkeley CA}}

\date{Submitted to Studies in Applied Math: February 24, 2005}
\begin{document}
\maketitle

\begin{abstract}
The problem of finding the optimal tapering of a free (non-supported) javelin is described and solved.
For the optimal javelin, the lowest mode of vibration has the highest possible frequency. 
With this tapering inner damping will lead to the cessation of the vibration at the fastest possible rate.
The javelin is modeled as a beam of uniform material.
The differential equations governing the vibration and the tapering of the beam are derived.
These equations have a difficult singularity at the tips of the beam.
A procedure using a similarity solution, as in
\cite{FN05:Tallest}, is used to solve this singular system, and the solution is found. The maximal frequency is found to be almost 5 times larger than the frequency of a cylindrical rod.
\end{abstract}


\section{Introduction}
The interest in the optimal design of columns, beams and plates has
existed for many years. Euler started the rigorous study of the
buckling load of columns and introduced the problem of designing the
strongest column. Keller, in 1960 solved this problem \cite{KellerStrongest}. In 1964 Keller and Niordson found the
design of the tallest self-weighted column \cite{KellerTallest}. Others have continued
studying the various qualities of bending rods and plates under various conditions.

In this paper we find the optimal design of a non-supported beam (picture an Olympic javelin in mid-air). 
The aim is to find the design whose lowest mode of vibration has the
largest frequency.
The optimal design is shown to have a frequency that is greater than that of a constant cross-section beam by a factor of 5.

To simplify the problem we make several working assumptions on the
permissible designs of the column.
The cross-sectional shapes at different points along the beam are assumed to be geometrically similar with fixed  orientation (see figure \ref{fig:geometry}).
Furthermore, we assume that the cross-sectional shape is convex.
The cross sectional area is allowed to vary throughout the length of
the beam (``tapering''). 
While maximizing the frequency, we hold the total volume of the beam
fixed.

Working within linearized theory, it is sufficient to consider standing waves confined to a single plane.
These standing waves and their temporal frequencies are solutions of an ODE eigenvalue problem.
The frequencies are functionals of the beam shape. 
This analysis seeks the tapering of a beam with fixed length and volume, which maximizes the lowest frequency.
Formally this is done by requiring that the frequency be stationary
with respect to variation of the beam tapering. 
This  gives an additional ODE which relates the tapering and the standing wave amplitude.

The ODE's and boundary conditions  form a closed system for
the tapering, standing wave amplitude and frequency of the optimal beam. 
Unfortunately, they are difficult to solve. 
Naive shooting methods fail to get close to the end of the beam and therefore do not allow for corrections of the initial conditions to be made. 
More sophisticated boundary value problem solvers also fail to converge.
In \cite{Niordson65} Niordson solved a similar problem by converting
the ODE to integral form and then performing an iteration which converges to the solution. 
This paper follows Niordson's paper loosely but since the boundary
conditions (BC) are different and the method of solution is different, we present the full derivation and solution here. 
Having different BC means that although this problem has the same
ODE's, the singularities at the tips are more severe in this case.
We use the same method of solution shown before in \cite{FN05:Tallest}.
First, in Section \ref{sec:setup} the equations that characterize the
optimal beam and the shape of vibration mode are found. As mentioned,
these equations are nonlinear and singular at the tips of the beam. 
In Section \ref{sec:solution} we reduce these singular equations to a
regular system of ODE's that can be easily solved using standard
numerical methods.
A similarity solution to the equations is found and used to ``peel
away'' the singularity at the tips. 
The resulting ODE's have a critical point and by starting near the
critical point on its stable manifold, the equations are solved
backwards numerically until the BC are satisfied. 
Since the stable manifold is two dimensional, a simple 1-parameter shooting algorithm employing a standard ODE integrator will determine the solution. 
\section{Derivation of the Boundary Value Problem}
\label{sec:setup}
\subsection{Setup}
Consider all possible beams, all of the same length and volume which
are suspended without gravity or other external forces. 
The beams have various modes of vibration. 
What is the design of the beam whose first vibration mode has the
largest frequency? 
To simplify, we solve the problem only for a specific class of
permissible designs. 
We assume that the beam is thin (i.e. the characteristic width is much
less than the length of the beam) and made of a homogeneous material. 
In addition, we only permit beams with geometrically similar, equally
oriented and convex cross-sections.
Lastly, we concern ourselves with the bending of the beam in a
specified plane only.

\begin{figure}[bh!]
\center{
\scalebox{0.4}{
\includegraphics{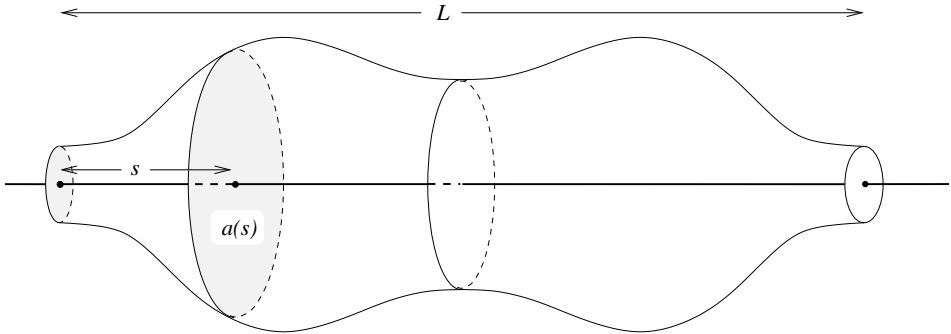}}}
\caption{The beam is assumed to be made of a homogeneous material with convex cross-section.
The area of the cross-section at point $s$ is denoted by $a(s)$. 
The cross-sections are geometrically similar and equally oriented. }
\label{fig:geometry}
\end{figure}

\begin{figure}[tbh!]
\center{
\scalebox{0.5}{
\includegraphics{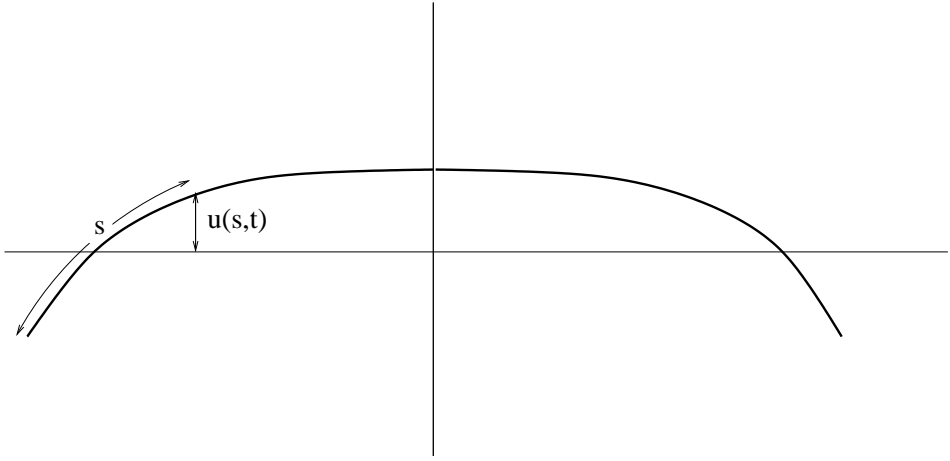}}}
\caption{A ``snapshot'' of the vibrating beam at time $t$.
The vertical displacement of the beam is given by $u(s,t)$.
The position along the beam is parameterized by arclength $s$, measured from one of the tips of the beam.}
\label{fig:variables}
\end{figure}

The beam is parameterized by arclength $s$, measured from one of the
tips along the beam's center axis.
The design information is contained in a single function $a(s)$, the
cross-sectional area of the beam at point $s$ (see Figure
\ref{fig:geometry}).  
At rest and under no stress, the center of beam is assumed to lie on the $x-$axis.
The beam configuration at time $t$ is specified by $u(s,t)$,  measuring the vertical displacement of the point $s$ from the $x-$axis (see Figure \ref{fig:variables}). 

First, we think of the cross sectional area, $a(s)$, as given.
The total volume of the beam is
\begin{equation}
  V[a] = \int_0^L a(s)\,ds,
\label{dim_volume}
\end{equation}
where $L$ is the total length of the beam.

\subsection{Lagrangian}
A beam design given by $a(s)$ determines the vibration modes of the
beam. 
To find the ODE that governs the vibration, we write the Lagrangian,
given by the difference between the kinetic and potential energy of the beam:
\begin{equation}
  \LL[u] = \int_0^L\BRK{-\half b(s)\,u_{ss}^2 + \half\rho\, a(s)\, u_t^2}\, ds.
\label{langrangian}
\end{equation}
Here, $\rho$ is the mass density of the material.
The function $b(s)$ is the bending modulus which is proportional to $a^2(s)$.
Specifically,
\begin{equation}
  b(s) = c\,E\,a^2(s),
\end{equation}
where $c$ is a dimensionless constant that depends on cross-section shape, and $E$ is the Young's modulus of the material.   
Using separation of variables we write the deflection function $u(s,t)$ as a product of a standing wave amplitude function, $y(s)$ and $\cos(\omega t)$, and average the Lagrangian \eqref{langrangian} over a temporal period:
\begin{align}
  \overline{\LL}[y] &=\frac{\omega}{2\pi}\int_0^{\frac{2\pi}{\omega}} \int_0^L\BRK{-\half c\,E\,a^2(s)\,y_{ss}^2 \cos^2(\omega t) + \half\rho\, a(s)\, y^2 \omega^2 \sin^2(\omega t)}\, ds\,dt.\\
&=\quarter \int_0^L\BRK{- c\,E\,a^2(s)\,y_{ss}^2  + \rho\, a(s)\, y^2 \omega^2}\, ds.
\label{average_lang}
\end{align}

The average Lagrangian can be written using non-dimensional variables by implementing the units in the scaling table:
\begin{center}
\begin{tabular}{l>{$}c<{$}>{$}c<{$}>{$}c<{$}>{$}c<{$}>{$}c<{$}}
Variable&a&s&y&\overline\LL\\
\hline
Unit&\frac {V}L&\frac L2&\sqrt{\frac {2V}L}&\frac{8cEV^3}{L^6}
\end{tabular}
\end{center}

The dimensionless versions of \eqref{average_lang} and \eqref{dim_volume} are
\begin{align}
  \overline\LL[y] &=\quarter \int_0^2\BRK{-a\,{y}_{ss}^2 + \lambda^2\, a\, y^2}\, ds,
\label{dimless_lang}\\
V[a]&\equiv\int_0^2a\,ds=2\label{dimless_vol}
\end{align}
Here, the square of the non-dimensional frequency is $\lambda^2=\frac{\rho L^4\omega^2}{16cEV}$.

\subsection{Vibration ODE}
The Euler equations of \eqref{dimless_lang} constitute a boundary value problem (BVP) for $y(s)$:
\begin{gather}
  \brk{a^2\,y_{ss}}_{ss} -\lambda^2 a y =0, \quad \text{in } 0<s<2,\\
  a^2y_{ss} =0  \text{ at } s = 0,2, \qquad (a^2y_{ss})_s=0 \text { at }s = 0,2.
\end{gather}
Physically, the BC express the absence of torque and force at the ends.
Although the Euler equations were derived from the average Lagrangian, finding the ODE from the full Lagrangian and then using separation of variables will lead to the same equations for $y(s)$.
 
Heuristically, it is reasonable to expect the optimal beam shape has $a(s)$ even.
We also expect the fundamental mode to be an even standing wave. 
This allows us to solve the problem on the interval $0\le s\le1$.
The volume constraint \eqref{dimless_vol} reduces to
\begin{equation}
  \int_0^1 a(s)\, ds =1.
\label{nondim_volume}
\end{equation}

At the endpoint $s=1$ we impose symmetry boundary conditions on $y(s)$, so the eigenvalue problem for the shape of standing waves is
\begin{equation}
  \brk{a^2\,y_{ss}}_{ss} -\lambda^2 a y =0,  \quad 0\le s\le1
\label{euler}
\end{equation}
\begin{align}
    a^2y_{ss}&=0, \text{ at } s=0, &  \brk{a^2y_{ss}}_s&=0, \text{ at } s=0,
\label{bc_tip}\\
  y_s(1)&=0,&  y_{sss}(1)&=0.\label{symmetry_bc}
\end{align}

\subsection{The Frequency of a Cylindrical Javelin}
\begin{figure}[ht!]
\center{
\scalebox{0.75}{\includegraphics{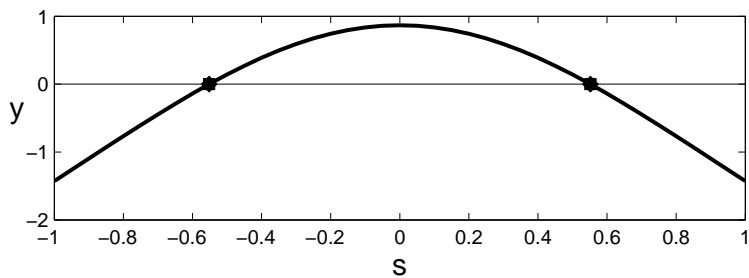}}}
\caption{The shape of the first vibration mode of the cylindrical javelin.}
\label{fig:cylinder_wave}
\end{figure}

If the cross-section $a(s)$ is a constant, the javelin is a simple cylinder. 
In this case we can solve the problem (almost) analytically. 
This will give us a reference frequency to compare with later. 
To find the frequency of a cylindrical javelin, it is more straightforward to shift the origin of $s$ and solve on the interval $\Brk{-1,1}$.
The even solutions to the ODE are

\begin{equation}
  y(s) =  A \cos\brk{\sqrt{\lambda}s}+  B \cosh\brk{\sqrt{\lambda}s}.
\end{equation}
The BC yield a constraint on $\lambda$:
\begin{equation}
  -\tan\brk{\sqrt{\lambda}} = \tanh\brk{\sqrt{\lambda}}
\end{equation}
Solving this equation numerically for the smallest (nonzero) $\lambda$ gives 
\begin{equation}
\lambda \approx 5.5933.
\end{equation}
This is the non-dimensional frequency of the cylindrical javelin.
The standing wave shape of the cylindrical javelin is shown in figure \ref{fig:cylinder_wave}

 \subsection{Maximizing the Frequency}
Up to this point, we found an eigenvalue problem that implicitly determines a functional, $\lambda[a]$.
For each cross-sectional area function $a(s)$ it defines the frequency of a standing wave solution of a beam tapered according to it. 
Although the formula is implicit and we can by no means give
an explicit formula for $\lambda[a]$, we would like to find the function $a(s)$ so $\lambda[a]$ is stationary with respect to variations of $a(s)$.
To do this, we find the functional derivative of $\lambda[a]$ (with respect to $a$) and write
\begin{equation}
\frac{\delta\lambda}{\delta a} = \mu \frac{\delta V}{\delta a}=\mu.
\label{func_der}
\end{equation}
In \eqref{func_der}, $V[a]$ is the volume functional \eqref{nondim_volume} and $\mu$ is a Lagrange multiplier associated with the volume constraint.
To find $\frac{\delta\lambda}{\delta a}$, we introduce a small variation to the cross-sectional area $\delta a(s)$.
Let $\delta \lambda$ and $\delta y$ be the resulting variations in $\lambda$ and $y$.
Assuming the the resulting variations are small when $\delta a$ is small, the linear variational equations which follow from (\ref{euler}--\ref{symmetry_bc}) are
\begin{align}
  (2a\,\delta a \,y_{ss} + a^2\,\delta y_{ss})_{ss} -
  2\lambda\,\delta\lambda\, a\,y - \lambda^2\,\delta a\,y
  -\lambda^2\,a\,\delta y =0,\label{variational}
\end{align}
\begin{align}
    2a\,\delta a\,y_{ss}+a^2\,\delta y_{ss}&=0, \text{ at } s=0, &  \brk{2a\,\delta a\,y_{ss}+a^2\,\delta y_{ss}}_s&=0, \text{ at } s=0,\\
  \delta y_s(1)&=0,&  \delta y_{sss}(1)&=0.
\end{align}
This is a linear, inhomogeneous BVP for $\delta y$. 
Its solvability condition determines the relationship between $\delta a$ and $\delta\lambda$ and thus gives $\frac{\delta\lambda}{\delta a}$.
The solvability condition is found by multiplying equation \eqref{variational} by $y$ and integrating from $0$ to $1$. 
Integration by parts, use of the BC, and rearrangement give
\begin{equation}
\int_0^1 \BRK{(a^2\,y_{ss})_{ss}-\lambda^2\,a\,y}\,\delta y\,ds + \int_0^1\BRK{ 2a\, y_{ss}^2  - \lambda^2\,y^2}\delta a \,ds = \delta\lambda\int_0^1  2\lambda \,a\,y^2   \,ds.
\end{equation}
The first integral in the left-hand integral vanishes due to \eqref{euler}. 
The remaining terms give the desired relationship between $\delta a$ and $\delta\lambda$:
\begin{equation}
\int_0^1 \brk{2a\, y_{ss}^2  - \lambda^2\,y^2}\delta a \,ds = \delta\lambda\int_0^1  2\lambda\, a\,y^2   \,ds,
\end{equation}
or
\begin{equation}
  \frac{\delta\lambda}{\delta a}= \frac{2a\, y_{ss}^2  - \lambda^2\,y^2}{2\lambda\int_0^1   a\,y^2   \,ds}.
\end{equation}
Substituting the expression for $\frac{\delta\lambda}{\delta a}$ into equation \eqref{func_der} yields an integro-differential equation that characterizes the optimal tapering $a(s)$,
\begin{equation}
2a\, y_{ss}^2  - \lambda^2\,y^2=2\mu\, \lambda\int_0^1 a\,y^2   \,dr  \label{optimal_beam}
\end{equation}
A short calculation shows that $\mu = \lambda$ (see Appendix \ref{mu_calc}).
Since the RHS of \eqref{optimal_beam}  is independent of $s$, this integral equation can be transformed into an ODE by differentiating it once with respect to $s$,
\begin{equation}
2\brk{a y_{ss}^2}_s  - \lambda^2\,\brk{y^2}_s=0.\label{optimal_diff}
\end{equation}
It would seem that in order to remain equivalent to the integro-differential equation \eqref{optimal_beam}, an additional BC should be added. 
In fact, the volume constraint on $a$ is enough. See Appendix \ref{mu_calc}.

The volume constraint \eqref{nondim_volume}, BVP
(\ref{euler}--\ref{symmetry_bc}) and equation \eqref{optimal_diff},
characterize the tapering of the javelin with highest frequency. 
These equations are singular at the tip, $s=0$, due to the BC \eqref{bc_tip} and a direct numerical approach fails to give a solution. 
We will now find a similarity solution that will remove the singularity by transforming the ODE into an autonomous system which can be solved using a simple ODE solver.
\section{Solution of the BVP}
\label{sec:solution}
In order to manage the derivatives more easily, we introduce a new variable,
\begin{equation}
  \phi = a^2\,y_{ss}.
\end{equation}
This is the non-dimensional torque.
%
Since the torque, $\phi$, goes to zero at the tip (due to the BC), and we expect $a(s)$ to go to zero as well, we look for an algebraic relation between the variables and the distance from the tip. 
The limit to be examined is $s\goto 0$. 
First we change the constraint $\int_0^1 a\,ds=1$ into a BVP.
This can be achieved by adding a variable $b(s)$:
\begin{equation}
b(s) =\int_0^s a(s')\,ds'.  
\label{b_def_int}
\end{equation}

In terms of the newly defined variables, the BVP is
\begin{align}
\phi - a^2\,y_{ss}&=0,\label{phi_def}\\
\phi_{ss} -\lambda^2\,a\,y&=0,\\
2\brk{\frac{\phi^2}{a^3}}_s -\lambda^2 \brk{y^2}_s&=0,\label{var_phi}\\
 b_s - a &= 0,\label{b_def}
\end{align}
\begin{align}
    \phi&=0, \text{ at } s=0, &  \phi_s&=0, \text{ at } s=0,
\label{bc_phi}\\
  y_s&=0, \text{ at } s = 1,&  a_s&=0, \text{ at } s = 1,
\label{new_symm}\\
b &=0, \text{ at } s=0, & b &=1, \text{ at } s=1.\label{b_bc}
\end{align}
In the equations above, BC \eqref{bc_phi} is the translation of \eqref{bc_tip}.
The BC \eqref{new_symm} are, in essence, symmetry BC. 
They follow from (\ref{symmetry_bc},\ref{optimal_diff}). 
Lastly, \eqref{b_bc} are the BC needed for the volume constraint. 
They follow immediatly from (\ref{nondim_volume}, \ref{b_def_int}).

\subsection{Similarity Solution}
The ODE's (\ref{phi_def}--\ref{b_def}) have a similarity solution which
satisfies the BC at the tip of the beam. 
(For information on similarity solutions see, for example,
\cite{Barenblatt} or \cite{Bluman}.)
To find it, we examine the scaling relations among the variables.
Let $A, B, P, Y$ and $S$ be the ``units'' of $a, b,\phi,y$ and $s$ respectively.
A balance of ``units'' in the equations (\ref{phi_def}--\ref{b_def}) gives the relationships
\begin{align*}
  P &= A^2YS^{-2},\\
  PS^{-2}&=AY,\\
  P^2A^{-3}S^{-1} &= Y^2S^{-1},\\
  BS^{-1} &= A.
\end{align*}
This system has a two-parameter family of solutions:
\begin{align*}
  A &= S^4,\\
  B &= S^5,\\
  P &= YS^6.
\end{align*}
This leads us to look for a solution (\ref{phi_def}--\ref{b_def}) of the form
\begin{align*}
  \hat a(s) &= a_0s^4,\\
  \hat b(s) &= b_0s^5,\\
\hat \phi(s) &= \phi_0s^{p+6},\\
\hat y(s) &= y_0s^p.
\end{align*}
In the equations above, the exponent $p$ is unknown.

Substituting these equations into (\ref{phi_def}--\ref{b_def})   yields
\begin{align}
  b_0 &= \frac{a_0}{5},\\
\phi_0 &= p(p-1)y_0\,a_0^2,\\
\gamma &= p(p-1)(p+6)(p+5),\label{gamma1}\\
\gamma &= 2p^2(p-1)^2,\label{gamma2}
\intertext{where}
\gamma &= \frac{\lambda^2}{a_0}.
\end{align}
Equations \eqref{gamma1} and \eqref{gamma2} yield a polynomial equation for $p$,
\begin{equation}
  p(p-1)(p+6)(p+5)= 2p^2(p-1)^2.\label{closed_indicial}
\end{equation}
Since we are looking for a real frequency, $\lambda^2$ must be positive.
Clearly, $a_0$ must also be positive since $a(s)$ is an area.
Therefore, $\gamma$ must also be positive. 
This rules out the two trivial solutions to \eqref{closed_indicial}, $p=0$ and $p=1$.
The two other solutions are $-2, 15$. 

The solution $p=15$ gives a vanishing LHS for  the integral equation \eqref{optimal_beam} (as $s\goto0$). 
This is not possible unless the constant RHS is also zero. 
Since the RHS of \eqref{optimal_beam} is  positive, $p=15$ is not a solution of interest.
This leaves us with the single possible solution $p=-2$.
This solution yields $\gamma=72$ and gives rise to the following similarity solution:
\begin{align}
  \hat a(s) &= \frac{\lambda^2}{72}s^4,\label{sim_a}\\
  \hat b(s) &= \frac{\lambda^2}{360}s^5,\\
  \hat \phi(s) &= y_0\frac{\lambda^4}{864} s^4,\\
  \hat y(s) &= y_0s^{-2}\label{sim_y}.
\end{align}
It is easy to check that (\ref{sim_a}--\ref{sim_y}) solves the ODE system for all $s$ and  satisfies the BC at the tip ($s=0$). 
It is also easy to check that this solution does not satisfy the BC at
the midpoint ($s=1$).
We now use this similarity solution to remove the singularity from the
ODE's, simplifying the equations to a point where a numerical solution
is possible.

\subsection{Peeling away the Singularity}
To analyze the solution of the full BVP (\ref{phi_def}--\ref{b_bc}), we ``peel away'' the similarity solution. 
This is done by the transformation to the variables $\alpha, \beta, \Phi, \zeta$ defined by
\begin{align}
   a(s) &= \hat a(s)\,\alpha(s),\label{ansatz1}\\
   b(s) &= \hat b(s)\,\beta(s),\\
   \phi(s) &= \hat \phi(s)\, \Phi(s),\\
   y(s) &= \hat y(s)\,\zeta(s).\label{ansatz4}
\end{align}
Substituting these expressions into the BVP (\ref{phi_def}--\ref{b_bc}) results in a BVP for $\alpha,\beta, \Phi, $ and $\zeta$. 
Since the resulting equations are homogeneous in $s$ we use $t=-\ln s$ as the independent variable.
In this variable, the ODE for $\alpha(t), \beta(t), \Phi(t), $ and $\zeta(t)$ are the autonomous system (AS),
\begin{align}
  6\Phi - \alpha^2(D+2)(D+3)\zeta &=0,\label{final_as1}\\
  (D-4)(D-3)\Phi - 12\alpha \zeta &=0,\label{final_as2}\\
(4+D)\frac{\Phi^2}{\alpha^3}-2\zeta(2+D)\zeta&=0,\label{final_as3}\\
(5-D)\beta -5 \alpha &=0\label{final_as4}.
\end{align}
Here $D$ is the derivative with respect to the variable $t$.
The BC for this system are:
\begin{align}
\Phi(t)e^{-4t} &\goto 0,& e^{-3t}(D-4)\Phi(t)&\goto 0,&\beta(t)e^{-5t} &\goto 0,&\text{ as } t&\goto\infty\label{final_BC1}\\
(D-4)\alpha &= 0,&(D+2)\zeta &=0,&\beta &=\frac{360}{\lambda^2}, &  \text{ at } t&=0. \label{final_BC2}\\
&\text{(a)}&&\text{(b)}&&\text{(c)}\notag
\end{align}
What have we gained by all these manipulations?
First, we notice that $\lambda$ is no longer part of the ODE. 
It only appears in the BC at $t=0$.
This greatly simplifies the solution of the BVP.
Also, we notice that the singularity at $s=0$ has been removed. 
The boundary conditions do not cause the variables to vanish and there is no delicate balance of terms. 
The similarity solution of the original BVP (\ref{sim_a}--\ref{sim_y})
is represented by the critical point
$(\alpha,\beta,\Phi,\zeta)\equiv(1,1,1,1)\equiv \one$.
Since the similarity solution satisfies the BC at the tip, we
look for a solution that satisfies the BC at t=0 and converges to $\one$ as
$t\goto\infty$. 
This means that we are looking for a solution on the stable manifold
of the fixed point $\one$.

As is shown in the next section, the stable manifold is
two-dimensional. 
On the other hand, the BC (\ref{final_BC2}a), (\ref{final_BC2}b) define a surface of co-dimension 2. 
Thus, these surfaces are expected to have discrete points of intersection. 
These points, via the BC (\ref{final_BC2}c)  determine a particular
value of $\lambda$. 
In our case we will find exactly one point and hence one possible
value for $\lambda$.
\subsection{Stable Manifold}
To find the tangent plane of the stable manifold of {\textbb 1}, we linearize the ODE around {\textbb 1} and search for
solutions of the form
\begin{equation*}
  y = y_0e^{qt}.
\end{equation*}
The directions with $\Re(q)<0$ are stable.
The linearization of the AS (\ref{final_as1}-\ref{final_as4}) is,
\begin{equation}
  \brk{\begin{array}{cccc}
-12&0&6 &-(2+D)(3+D)\\
 -12&0&(D-4)(D-3)&-12\\
-3(4+D)&0& 2(4+D)&-2(4+D)\\
-5&(5-D)&0&0

  \end{array}
}
\brk{\begin{array}{c}
  \delta\alpha\\
  \delta\beta\\
  \delta\Phi\\
  \delta\zeta
\end{array}}
=0.
\label{lin_as}
\end{equation}
Here, $(\delta\alpha,\delta\beta,\delta\Phi,\delta\zeta)$ are
deviations from {\textbb 1}.
We look for solutions of this system in the form
\begin{equation*}
  (\delta\alpha,\delta\beta,\delta\Phi,\delta\zeta) = (\delta\alpha_0,\delta\beta_0,\delta\Phi_0,\delta\zeta_0) e^{qt}.
\end{equation*}
Substitution  into \eqref{lin_as} yields 
\begin{equation}
  \brk{\begin{array}{cccc}
-12&0& 6 &-(2+q)(3+q)\\
 -12&0&(q-4)(q-3)&-12\\
-3(4+q)&0&2(4+q)&-2(4+q)\\
-5&(5-q)& 0&0
  \end{array}
}
\brk{\begin{array}{c}
  \delta\alpha_0\\
  \delta\beta_0\\
  \delta\Phi_0\\
  \delta\zeta_0
\end{array}}
=0.
\label{lin_sys}
\end{equation}
This system has a nonzero solution for
$(\delta\alpha_0,\delta\beta_0,\delta\Phi_0,\delta\zeta_0)$ when the
matrix in \eqref{lin_sys} is singular.
This happens for 6 values of $q: q_1 = 0,$ $q_2 = 1,$ $q_3 =-4,$ $q_4 = 5,$ $q_5 \approx 6.3523,$ $q_6\approx -5.3523$.
 The corresponding solutions for
 $(\delta\alpha_0,\delta\beta_0,\delta\Phi_0,\delta\zeta_0)$ are given
 in Table \ref{tab:eigenvalues}.
\begin{table}[hbt!]
 \begin{center}
   \begin{tabular}{>{$}c<{$}|>{$}c<{$}>{$}c<{$}>{$}c<{$}>{$}c<{$}>{$}c<{$}>{$}c<{$}}
     &S_1&S_2&S_3&S_4&S_5&S_6\\
     \hline
     \delta\alpha_0&0&4&9&0&0.68568&-11.019\\
     \delta\beta_0&0&5&5&1&-2.5352&-5.3220\\
     \delta\Phi_0&1&4&-27&0&1&1\\
     \delta\zeta_0&1&-2&135&0&-0.028525&17.529
   \end{tabular}
\caption{The stable and unstable directions around the critical point
  of the AS (\ref{final_as1}-\ref{final_as4}).}
\label{tab:eigenvalues}
 \end{center}
\end{table}

The values for $S_5$ and $S_6$ are approximate.
The only two stable solutions are $(q_3,S_3),$ $(q_6,S_6)$; therefore,
the plane tangent to the stable manifold is spanned by the two vectors $S_3$ and $S_6$.
The unstable direction $S_2$ is due to the similarity solution (see
\cite{FN05:Tallest}), and the unstable direction $S_4$ is due to
$\beta$ representing an integral constraint on the solution, not
actually  coupled to the ODE (See Appendix \ref{app:unstable}.)

To find the numerical solution of the BVP, we start near the fixed point $P$, on the plane tangent to the stable manifold, and solve the AS (\ref{final_as1}--\ref{final_as4}) backwards in $t$. 
The stopping condition is that both BC (\ref{final_BC2}a)  and  (\ref{final_BC2}b) are satisfied  at the same $t$. 
Since the stable manifold is two-dimensional, we have a one-parameter family of solutions each starting at a different direction on the manifold.
We use the shooting method to find the initial direction, so the
resulting solution for $(\alpha,\beta,\Phi,\zeta)$ satisfies both BC
at the same $t$. 
Since the system is autonomous, we redefine this $t$ to be zero.
Once stopped, the value of $\lambda$ will be determined from (\ref{final_BC2}c) and then the full solution follows using (\ref{ansatz1}--\ref{ansatz4}).
\section{Numerical Results}
Here is the actual mechanism of the shooting method:
First, the direction in the stable manifold is defined using a parameter $\theta\in\brkco{0,2\pi}$:
\begin{equation}
v(\theta) = \sin(\theta) S_3 + \cos( \theta) S_6.
\end{equation}
Next, the AS is solved backwards in $t$ starting from 
\begin{equation}
 x_0= (1,1,1,1)+\eps v(\theta),
\end{equation}
where $\eps$ is a small parameter determining how close to the fixed point to start the solution. 
A value of $\eps = 0.001$ was used in this numerical solution.
As the ODE is 6th order, initial values for $\zeta_t$ and $\Phi_t$
are needed. 
For this we used the derivatives of the similarity solution: $\zeta_t=\eps\, q\, \delta\zeta_0$ and  $\Phi_t=\eps\, q\,
\delta\Phi_0$.

\begin{figure}[th!]
  \center{
\scalebox{0.75}{
\includegraphics{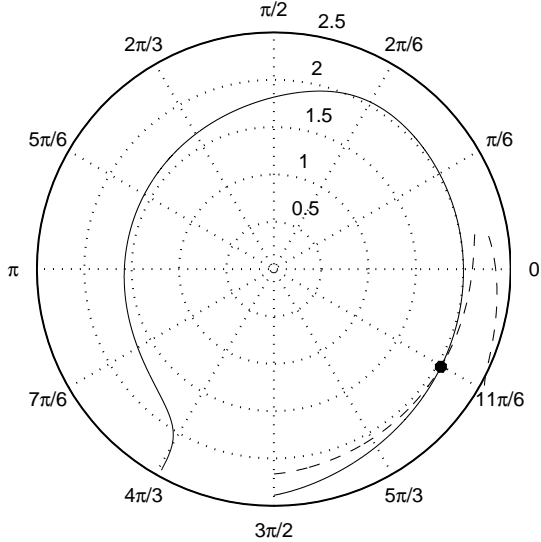}
}}
\caption{
The $-\Delta t$ at which each of the relevant BC are satisfied as a function of $\theta$.
The continuous line is the value of $-\Delta t$ for which $a_s=0$, and
the broken line is the value for which $y_s=0$. 
The plot shows that not every direction leads to a nicely behaved
solution. 
In some directions the solution explodes before one or both of the BC are satisfied.
In others the solution has two values of $t$ for which a particular BC is satisfied.}
\label{fig:events}
\end{figure}

\begin{figure}
  \center{
\scalebox{0.75}{
\includegraphics{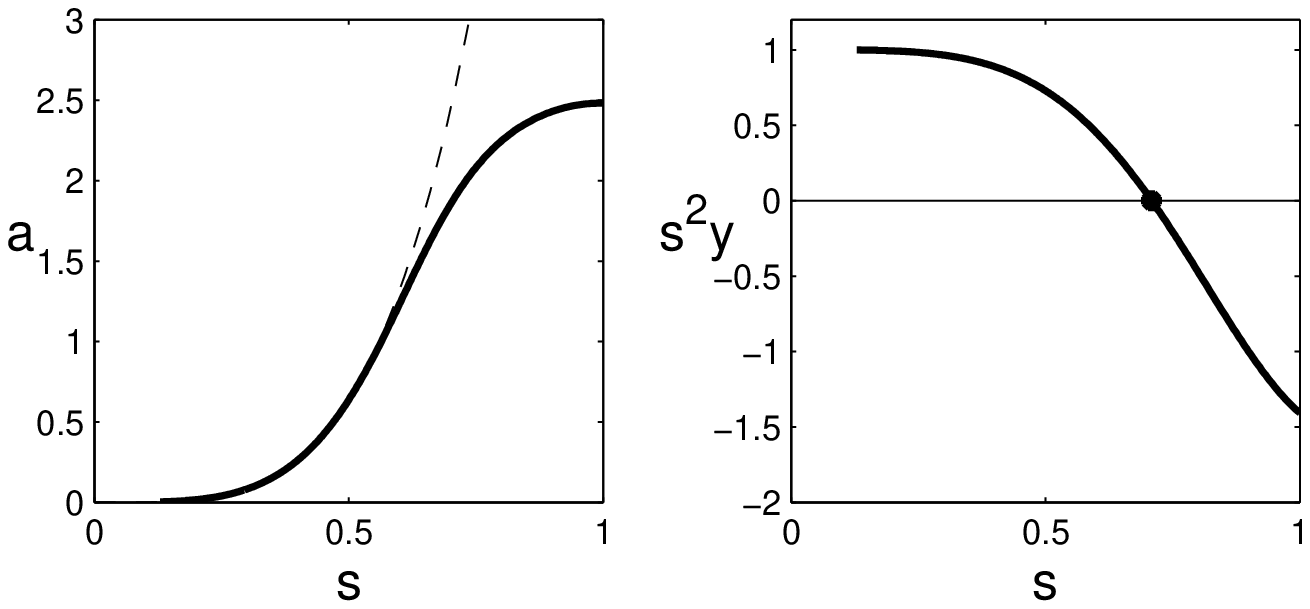}}}
$\begin{array}{c@{\hspace{2in}}c}
\mbox{(a)}&\mbox{(b)}
\end{array}$
\caption{The shape of the stiffest rod and the shape of the fundamental standing wave.
Since the fundamental standing wave has an $s^{-2} $ singularity at the tip, $s^2y$ is plotted instead of $y$.
Both plots do not continue all the way to $s=0$. 
This is because the solution was started a small distance away from the fixed point $P$. 
The solution can be easily continued near $s=0$ using the similarity solution.
The dashed line in the figure on the left is the similarity solution
$\tilde a(s)$, the actual solution slowly leaves this solution as $s$ increases.}
\label{final_solution}
\end{figure}

The AS is solved using \MATLAB ode solver \verb!ode45! using default
tolerances.
Plotting $-\Delta t$ for which each of the two BC are satisfied (for
each value of $\theta$), gives Figure  \ref{fig:events}. 
We see that for some values of $\theta$ one or both of the BC are
never satisfied, while for others a BC can be satisfied several
times. 
The two BC are satisfied for the same $-\Delta t$ for a single value
of $\theta$, around $\pi/6$.
Using the \MATLAB non-linear solver, \verb!fzero!, the value of $\theta$
where the two BC are satisfied at the same  $t$ is found,
$\theta\approx 5.753$. The solver was given $-\pi/6$ as the initial
guess for $\theta$.
For this $\theta$ the two BC are satisfied at $\Delta t = -2.0429$. 
From the value of $\beta$ at $s=1$ the value for $\lambda$ is found:
$\lambda\approx 27.073$. 
We can compare this value of $\lambda$ to the value for the simple cylinder.
The non-dimensional frequency for a cylindrical rod is 5.5933, and therefore the optimized rod vibrates almost 5 times faster than the cylindrical one.
The tapering of the optimal javelin is shown in Figure \ref{final_solution} along with the shape of its standing wave.
\section{Discussion}
We have shown how to use the similarity solution to  remove the singularity from the differential equations and find a solution that would otherwise require an iterative method.
 The variational equations were derived under the assumption that the spectrum of the differential operator (\ref{euler}--\ref{symmetry_bc}) is discrete and therefore the variation will have a meaning. 
Cox and McCarthy have shown (for example in \cite{CoxMcCarthy}, \cite{McCarthy}) that this is not always the case and that special treatment due to the existence of a continuous spectrum may be necessary. 
The existence of the continuous spectrum is due to the singularly tapered tips and therefore any minimal amount of rounding of the tips will eliminate the continuous spectrum. 

Another possible inaccuracy in the above derivation is due to the basic assumption that the deflection is small. 
The deflection $y$ ends up having a singularity at the tips of the beam and therefore can only be small away from the tips. 
This means that the linearization is a crude estimate at the tips. 
In addition the curvature was taken to be equal to $y_{ss}$, this is only true when $y_s\ll1$. 
Again, this assumption breaks down near the tips where the slope, $y_s$,  tends to infinity.

The extension of this analysis, to optimizing higher modes, is not obvious.
The second mode is expected to be anti-symmetric and can be found using other BC at the middle of the beam ($y=0$ instead of $y_s=0$).
Higher modes may have singularities at internal points. 
To solve this ``contact conditions'' governing the internal
singularities must be derived and used to connect between different parts of the solution.

\appendix
\section{Appendix}
\subsection*{Calculation of the Lagrange Multiplier $\mu$}
\label{mu_calc}
To calculate the Lagrange multiplier in \eqref{optimal_beam}, we multiply the equation by $a$, integrate and use the volume constraint \eqref{nondim_volume}
\begin{equation}
  2\int_0^1 a^2\,y_{ss}^2\,ds-\lambda^2\int_0^1\,a\,y^2\, ds = \mu\,\lambda\int_0^1 a\,y^2 \,dr.
\end{equation}
Two integration by parts (and use of the BC) yields
\begin{equation}
  2\int_0^1 \brk{a^2\,y_{ss}}_{ss}\,y\,ds-\lambda^2\int_0^1 \,a\,y^2\, ds = \mu\,\lambda\int_0^1 a\,y^2 \,dr.
\end{equation}
Using the ODE \eqref{euler} we get
\begin{equation}
  2\lambda^2\int_0^1 a\,y^2\,ds-\lambda^2\int_0^1 a\,y^2\, ds = \mu\,\lambda\int_0^1 a\,y^2 \,dr.
\end{equation}
Thus  $\mu = \lambda$.
By this calculation one can also ``go back'' from the differential equation \eqref{optimal_diff} to the integro-differential equation \eqref{optimal_beam}.
Integrating  \eqref{optimal_diff} once gives
\begin{equation}
  2\int_0^1 a^2\,y_{ss}^2\,ds-\lambda^2\int_0^1\,a\,y^2\, ds = C.
\end{equation}
Here $C$ is an unknown constant.
Multiplying by $a$, integrating and using the volume constraint \eqref{nondim_volume} recovers the constant $C$.

\subsection{An Unstable Direction}
\label{app:unstable}
Solution $S_4$ in table \eqref{tab:eigenvalues} has a suspicious form. 
The eigenvalue is 5, the exponent of the similarity solution, and the
eigenvector has components only in $\beta$ direction. 
This is because the original ODE are invariant under a shift of $b$ by
an additive constant. 
Shifting $\hat b$ by $\eps$ translates to the $\beta$ variable:
\begin{align*}
  \beta &= \frac{\hat b+\eps}{\hat b}\\
  &= 1+ \eps \frac{360}{\lambda^2}s^{-5}\\
  &= 1+ \eps \frac{360}{\lambda^2}e^{5t}.
\end{align*}
So we see that there is an unstable direction about the critical point
that makes $\beta$ increase exponentially with constant $5$.

\bibliographystyle{amsplain}

\bibliography{general}

\providecommand{\bysame}{\leavevmode\hbox to3em{\hrulefill}\thinspace}
\providecommand{\MR}{\relax\ifhmode\unskip\space\fi MR }
\providecommand{\MRhref}[2]{%
  \href{http://www.ams.org/mathscinet-getitem?mr=#1}{#2}
}
\providecommand{\href}[2]{#2}
\begin{thebibliography}{1}

\bibitem{Barenblatt}
G.~I. Barenblatt, \emph{Scaling, self-similarity, and intermediate
  asymptotics}, Cambridge University Press, New York, 1996.

\bibitem{Bluman}
G.~W. Bluman and S.~C. Anco, \emph{Symmetry and integration methods for
  differential equations}, Applied mathematical sciences, vol. 154,
  Springer-Verlag, 2002.

\bibitem{CoxMcCarthy}
Steven~J. Cox and C~Maeve McCarthy, \emph{The shape of the tallest column},
  SIAM Journal of Applied Math Anals \textbf{29} (1998), no.~3, 547--554.

\bibitem{FN05:Tallest}
Y.~Farjoun and J.~Neu, \emph{The tallest column --- a dynamical system approach
  using a symmetry solution}, Studies in Applied Mathematics \textbf{115}
  (2005), 319--337.

\bibitem{KellerStrongest}
J.~B. Keller, \emph{The shape of the strongest column}, Arch. Rational Mech.
  Anal \textbf{5} (1960), 275--285.

\bibitem{KellerTallest}
J.~B. Keller and F.~I. Niordson, \emph{The tallest column}, J. Math. Mech.
  \textbf{16} (1966), 433--446.

\bibitem{McCarthy}
C.~Maeve McCarthy, \emph{The tallest column --- optimality revisited}, Journal
  of computational and applied mathematics (1999), no.~101, 27--37.

\bibitem{Niordson65}
Frithiof~I. Niordson, \emph{On the optimal design of a vibrating beam}, Optimal
  Design \textbf{XXIII} (1965), no.~1, 47--53.

\end{thebibliography}

\end{document}